\documentclass{article}
\usepackage[utf8]{inputenc}
\usepackage{tabularx} 
\usepackage{amsmath}  
\usepackage{listings}
\usepackage{amssymb}
\usepackage{graphicx} 
\usepackage[margin=1in,letterpaper]{geometry} 
\usepackage[final]{hyperref} 
\hypersetup{
	colorlinks=true,       
	linkcolor=blue,        
	citecolor=blue,        
	filecolor=magenta,     
	urlcolor=blue         
}
\usepackage[style=numeric-comp,sorting=none]{biblatex}
\usepackage{graphicx,caption}
\usepackage{subcaption}
\usepackage{xcolor}
\definecolor{codegreen}{rgb}{0,0.6,0}
\definecolor{codegray}{rgb}{0.5,0.5,0.5}
\definecolor{codepurple}{rgb}{0.58,0,0.82}
\definecolor{backcolour}{rgb}{0.95,0.95,0.92}
\lstdefinestyle{mystyle}{
    language=C++,                		
	numbers=left,                  			%
	stepnumber=1,                   			
	numbersep=10pt,                  		
	backgroundcolor=\color{white},  
	commentstyle=\color{Blue},
    keywordstyle=\color{RubineRed},
    numberstyle=\tiny\color{codegray},
    stringstyle=\color{codepurple},
	showspaces=false,               		%
	showstringspaces=false,         		%
	showtabs=false,                 			
	frame=single,	                			
	breaklines=true,                		
	breakatwhitespace=false,        		
	escapeinside={\%*}{*)}     
}
\usepackage{blindtext}
\usepackage{textgreek}
\usepackage{amsmath}
\usepackage{gensymb}
\usepackage{float}
\usepackage{romannum}
\usepackage{tikz}
\usepackage{caption}
\usepackage{subcaption}
\usepackage{hyperref}
\usepackage{breakcites}
\usepackage{url} 
\usepackage{hyperref}
\usepackage{amsmath}
\usepackage{braket}
\usepackage{physics} 
\usepackage{mathtools}
\usepackage[utf8]{inputenc}
\usepackage{array, booktabs}
\usepackage{array}
\usepackage{graphicx}
\usepackage{caption}

\usepackage{multirow} 
\usepackage{amssymb} 
\usepackage{graphicx} 
\usepackage{array} 
\usepackage[labelfont=bf]{caption}

\usepackage{authblk}
\usepackage{svg}
\usepackage{amsmath}

\begin{document}
\pagenumbering{arabic}
\title{Exploring the origin of stronger survival of polarized vortex beams through scattering media}
\author[1*]{Atharva Paranjape}

\author[1*+]{Shyamal Guchhait}

\author[2]{Athira B S}

\author[1,3]{Nirmalya Ghosh}
\affil[1]{Department of Physical Sciences\\

Indian Institute of Science Education and Research Kolkata\\

Mohanpur, India - 741246}
\affil[2]{Department of Bioengineering\\
University of Washington\\
Seattle, USA - 98105}

\affil[3]{Centre of Excellence in Space Sciences India\\

Indian Institute of Science Education and Research Kolkata\\

Mohanpur, India- 741246}

\affil[+]{\href{mailto:sg16ip022@iiserkol.ac.in}{sg16ip022@iiserkol.ac.in}}
\affil[*]{these authors have contributed equally to this work}

\date{}
\maketitle


\begin{abstract}
Laguerre-Gaussian (LG) beams carrying orbital angular momentum (OAM) have shown promise in deep tissue imaging, medical diagnostics, and optical communication due to their robust propagation properties through scattering media. However, an exact model that provides a mechanism for the enhanced scattering properties of LG beams over Gaussian beams has not been established till date. Here, we examine this issue by studying the propagation of polarized vortex beams transmitted through tissue-like turbid scattering media. We demonstrate that the intensity profile has a much more profound effect on depolarization than the phase profile for LG beams. Our results indicate that the observed stronger propagation for the higher order LG beams is due to a higher anisotropy factor g as seen by the incident beam. This insight is expected to contribute towards building a complete picture of light transport in the presence of scattering, as well as guide optimization of the intensity and polarization structure of light for use in biomedical applications. 
\end{abstract}


\section{Introduction}
\
Since its first direct observation in 1992 \cite{allenoam}, light carrying orbital angular momentum (OAM) has invoked several studies in an endeavour to utilize its unique properties. Some of the applications include trapping and transfer of OAM to microscopic particles, encoding information through spatial modes as well as spin-orbit photonics \cite{oamtransfer,OAMtrapping, oamcommunication,soi,athira_josa}. In the field of biomedical optics, OAM has already been used for endoscopy, optical biosensors, microscopy, optical diagnostics, tissue optics and spectroscopy \cite{mousebrain,biomedical}. Spatial modes of light have opened up new avenues in classical and quantum communication and propagation of OAM modes has been explored in free space, optical fibres and underwater \cite{freespace,underwatercomm}. This promising degree of freedom of light offers a theoretically infinite dimensional Hilbert space, strengthening its potential for quantum cryptography \cite{OAMcryptography}. Non-local quantum entanglement of OAM modes has also been demonstrated and recently extended to higher dimensions
\cite{OAMentanglement,highoamentangle}.

A common limitation faced across the range of applications is the loss of phase structure, degradation of the spatial profile and depolarization of light due to scattering during propagation through various media. A detailed understanding of how structured beams undergo scattering will allow tailoring of the beam modes for specific applications and make full use of the versatile polarization and spatial modes that light offers. This has motivated various investigations of the scattering properties of light in an effort to make these technologies more resilient \cite{tissuephantoms}. As a particular class of OAM carrying light, the use of Laguerre-Gaussian (LG) beams has been widespread. Wang et al. \cite{Deeptransmission} have demonstrated that LG beams show greater depth penetration through scattering media and that the survival of the LG beams has a strong dependence on the topological charge \cite{Deeptransmission}. Recently, the dependence of transmission on the mode order has also been demonstrated in biological scattering material using chicken breast tissue \cite{chickentissue}. These studies consider the transmittance of light in different scattering regimes and show that higher-order LG beams have higher transmittance than Gaussian beams in the multiple-scattering regime. 

The mechanism of scattering of plane waves by dielectrics has been well established within the framework of Mie theory \cite{mie}. The dependence of polarization and the transport of polarized light through turbid scattering media has also been studied extensively through experiment \cite{GhoshVitkin} and numerical simulations \cite{MCxu,jessicaMC}. Thus the depolarization behaviour can be used to extract information about the scattering dynamics. These models, however, have not been extended for beams with more complex intensity and phase patterns such as the LG beams yet. Hence a model informed by experiments that explains the scattering differences between structured light and gaussian beams is desired and invites further investigation. Recently several efforts have been made in this direction. Gianani et al. \cite{vvbDOP} have studied the propagation of vector vortex beams through scattering media and carried out studies on depolarization and spatial profile degradation of the beams. Li et al. \cite{Ince} have showed that Ince-Gaussian (IG) vector beams have longer survival than scalar IG beams and Gaussian beams using depolarization analysis.

In this work, we investigate the origin of the scattering differences between fundamental Gaussian and LG beams using a simple experimental setup. With polystyrene latex bead solutions as  turbid scattering media, we collect the forward scattered light and perform polarization measurements. For a range of concentrations and varying particle sizes, we characterize the dependence of the degree of polarization (DOP) of the collected light on the topological charge of the input LG beam. We then proceed to isolate the effect of the phase structure from the increasing ring radius of LG beams, using the recently introduced perfect vortex beams, \cite{pvoriginal} whose radial size remains constant with changing mode order. In this way, we extend the previous analysis on propagation of structured light through scattering media. We present data for bead size of $3  \mu m$ and $0.1 \mu m$, which represent large and small sizes as compared to the wavelength of the scattered light respectively. Our observations indicate that the scattering differences originate predominantly from the varying intensity profile and not from the phase structure. We also propose that the mechanism of stronger propagation of higher-order LG
beams is through a varying anisotropy parameter g as seen by the beam.


The rest of the paper is organized as follows. In section \ref{section2} we outline our experimental setup and provide details of the parameters used and varied during the experiment. In section \ref{section3} we present our results and analysis. We conclude in section \ref{section4} and provide an outlook.


\section{Experiment}
\label{section2}
\begin{figure}[ht]
    \centering
    \includegraphics[width=\columnwidth]{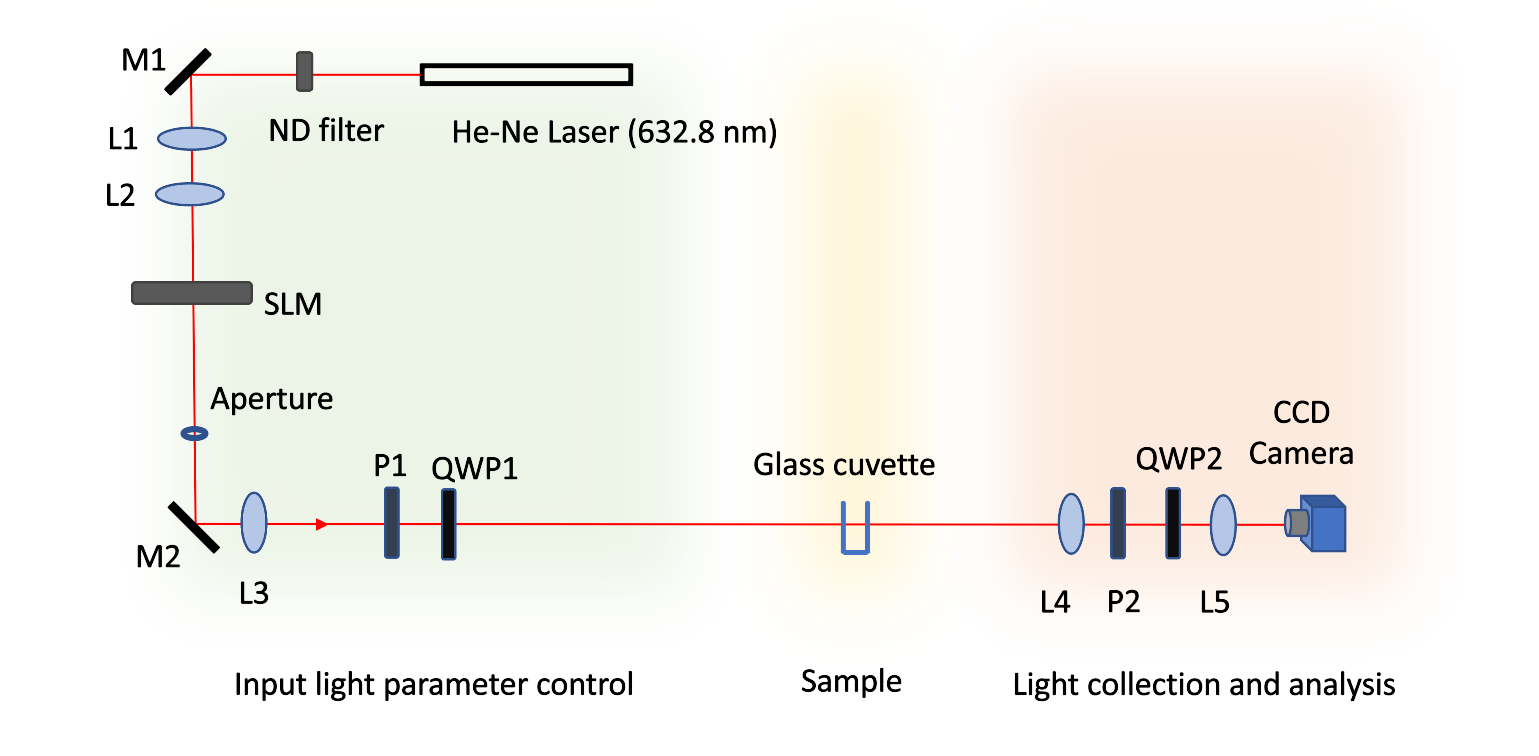}
    
    \caption{ \textbf{Schematic of the experimental setup}. The coloured sections distinguish various parts of the experiment: Input light parameter control (Green), Sample (Yellow), Light collection and analysis (Pink). (M1, M2) Mirrors, (L1, L2, L3, L4, L5) Lens, (P1, P2) Polarizer (QWP1, QWP2) Quarter wave plate,  SLM (Spatial Light Modulator), ND (Neutral Density) Filter. Light with required OAM value and polarization is generated using an SLM and a set of polarizer (P1) and quarter wave plate (QWP1). This light is then made incident on the cuvette containing a sample of scattering media. Polarization measurements are performed on the forward scattered light using another set of polarizer(P2) and quarter-wave-plate (QWP2) and a CCD camera.}
    \label{fig:setup}
\end{figure}

The schematic of the experimental setup is shown in Fig.\ref{fig:setup} and consists of a He-Ne laser (Thorlabs) with a working wavelength of 632.8 nm. 
Neutral density filters are used to control the beam intensity as required during the experiment. A combination of two lenses (L1, L2) is used to control the beam size and focus. A slightly focused beam is made incident on a spatial light modulator (SLM, Holoeye, LC2012). The SLM is used to modify the phase of the input beam while the amplitude is kept unaltered. Computer-generated fork hologram patterns are projected to the SLM to generate the required beams in the transmission mode output. Forked hologram patterns corresponding to topological charge values ranging from $l$ = 0 to 10 are used in the experiment. Multiple orders of the LG beams are generated by the SLM, which can be resolved at the effective focal point of the lens combination. An aperture is used to select the required order and subsequently a lens (L3) is used to collimate the beam. A polarizer (P1) and quarter wave plate(QWP1) are then used to create horizontally (H) or left circularly polarized (LCP) input light. 

To perform polarization measurements, we collect the diffused as well as ballistic light immediately after the sample. The ballistic light which has not undergone multiple scattering retains most of its polarization, while the multiple scattered diffuse light undergoes significant depolarization. When the analyzer (P2) is in a crossed configuration with the polarizer (P1), the ballistic light is cut off and we observe the diffused light prominently (See supplementary Sec S.3 ). To analyze the complete polarization state, a set of six measurements are carried out by rotating the polarizer (P2) and quarter wave plate (QWP2) as required for the Stokes parameter measurement. For this measurement, we use the horizontal, vertical (H, V), diagonal, anti-diagonal (D, A), right and left circularly polarization (R, L) bases and the corresponding projected intensities, $I_{H}, I_{V}, I_{D}, I_{A}, I_{R}, I_{L}$. The complete polarization state is given by the Stokes vector $\overrightarrow{S}$ as  

\begin{equation}
\centering
   \overrightarrow{S}  =   \begin{pmatrix}
    I \\
    Q \\
    U \\
    V \\
\end{pmatrix}  
\end{equation}

\noindent where, 
\begin{equation}
    Q = \frac{I_{H}-I_{V}}{I_{H}+I_{V}} , \hspace{1 cm} U = \frac{I_{D}-I_{A}}{I_{D}+I_{A}} , \hspace{1 cm}  V = \frac{I_{R}-I_{L}}{I_{R}+I_{L}}
\end{equation}

\noindent Hence the degree of polarization is given as, 
\begin{equation}
   DOP = \sqrt{Q^2 + U^2 + V^2}
\end{equation}

From the set of six measurements, the pixel-wise degree of polarization is calculated. In all instances, we report the degree of polarization as defined above.


The path length of the light through the sample is kept at 1 mm throughout the experiment. Polystyrene microsphere bead solutions (Sigma-Aldrich) are used to prepare the scattering media. The stock solutions are appropriately diluted using 0.1 \% w/v Sodium Dodecyl Sulphate (SDS) solution prepared using distilled water. The particle sizes and scattering coefficients of the prepared samples are summarized in Table \ref{tab:parameter_table}. 


\begin{table}[H]
    \centering
    \begin{tabular}{|c|c|c|c|c|}
    \hline
    Particle diameter, \textit{d} ($\mu m$)  &   $g$    &    $\mu_{s} ( mm^{-1})$   &   $\mu_{s}' ( mm^{-1})$  &   $\tau = \mu_s\times d$  \\                          \hline
    && 1.1   &   1   &   1.1    \\
    0.1 & 0.091 &  3.302   &   3   &   3.302   \\
    &&  5.503   &   5   &   5.503   \\
    &&  11.006  &   10  &   11.006  \\
    \hline
    && 5.401    &   1   &   5.401   \\
    3   & 0.814   &  16.204  &   3   &   16.204  \\
    &&  27.007  &   5   &   27.007  \\
    && 54.014   &   10  &   54.014\\
    \hline
\end{tabular}
\caption{Set of sample parameters used in the experiment }
\label{tab:parameter_table}
\end{table}

 The reduced scattering coefficient is given by $\mu_{s}'$ = $\mu_{s}$ $\cross$(1 - g) where g is the scattering anisotropy parameter and $\mu_{s}$ is the scattering coefficient. The chosen $\mu_s'$ values allow investigation of the multiple scattering regime where the differences between LG beams and fundamental Gaussian beams are expected to be pronounced. 

In the second part of the experiment, we extend the same setup to generate perfect vortices following the method proposed by Jabir et al. \cite{pvgeneration}. Perfect vortex beams are generated by using the LG beam as an input to a pair of axicon lens and a biconvex lens. By varying the distance between the axicon lens and the Fourier lens, a perfect vortex beam of appropriate size is generated at the focal point of the Fourier lens. The sample is then kept at the focal point and the forward scattered light is collected immediately after the sample. The lens combination propagates and focuses the perfect vortex onto the camera. The polarizer (P2) and quarter wave plate (QWP2) are placed in the light path just before the camera and used to measure the pixel-wise DOP.


\section{Results and Discussion}
\label{section3}

\begin{figure}[H]
\centering
\includegraphics[width=.75\linewidth]{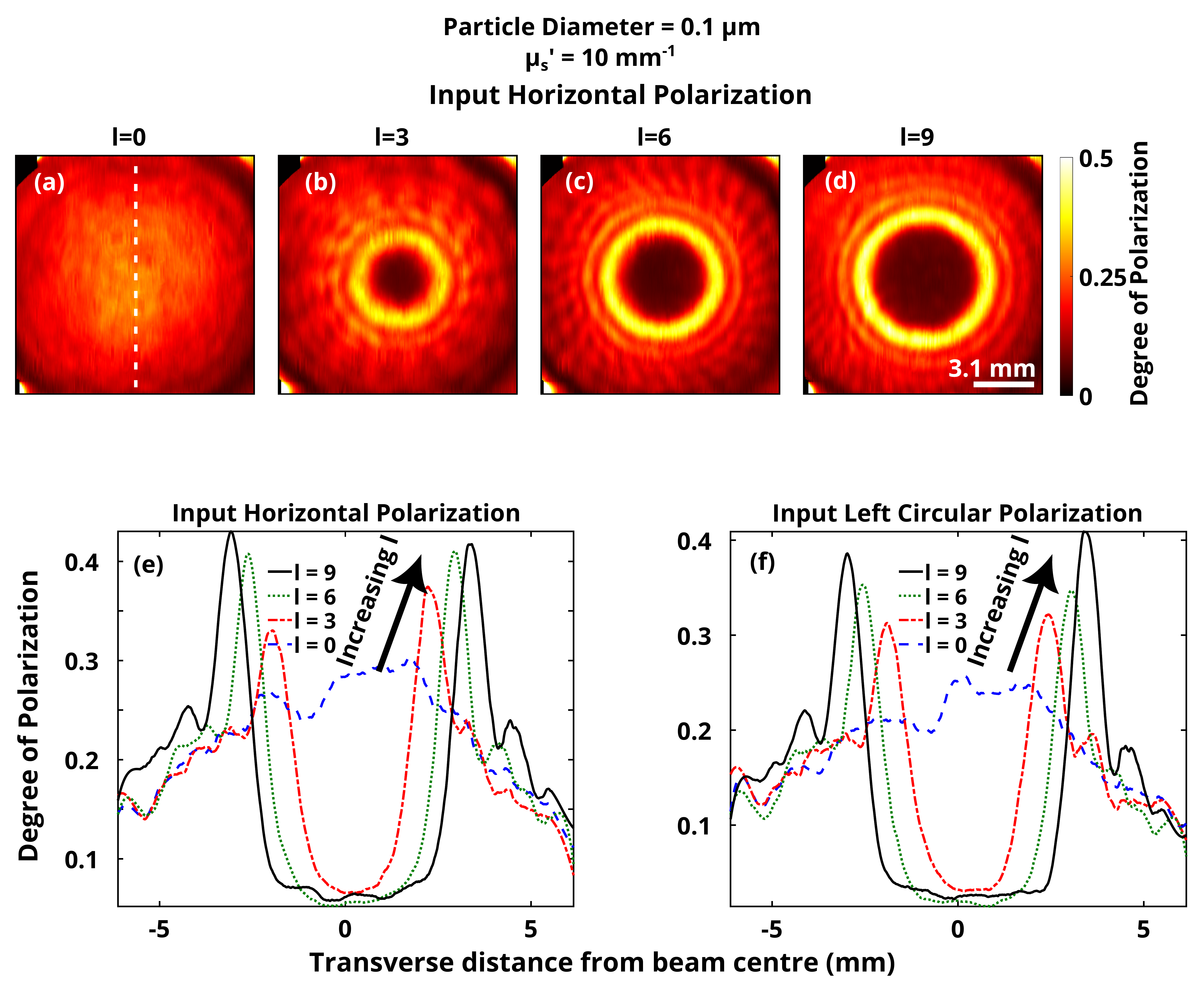}
\caption{\textbf{Influence of the OAM topological charge $l$ on the propagation of polarized light and the resultant degree of polarization (DOP) of the transmitted light in isotropic scattering media$(g = 0.091, d= 0.1\ \mu m, \mu_{s}' = 10\ mm^{-1} )$}.\textbf{((a)-(d))} Measured pixel-wise DOP in transmitted LG beam for incident light carrying topological charge $l$ = 0, 3, 6, 9 respectively. The colourbar represents the value of the degree of polarization. \textbf{(e)} and \textbf{(f)} DOP variation along a segment selected parallel to the y-axis in the pixel-wise DOP plot for incident horizontally and left circularly polarized light respectively. Representative segment is shown in  \textbf{(a)}. DOP values are shown for $l$ = 0 (Blue dashed line), $l$ = 3 (Red dash-dotted line), $l$ = 6 (Green dotted line), $l$ = 9 (Black solid line). Input light with a higher $l$ value is seen to exhibit stronger propogation and  preservation of polarization for both linearly and circularly polarized light. For a fixed value of $l$, depolarization of circularly polarized light is observed to be stronger than horizontally polarized light. 
}
\label{fig:particle_a}
\end{figure}

We first consider the effect of the varying OAM values in the input LG beam on the measured DOP in the transmitted beam. Fig.\ref{fig:particle_a}(a)-(d)  shows the measured pixel-wise DOP values in the transmitted beam in the form of 2D plots for $l$ = 0, 3, 6, 9 respectively.  The 2D plots correspond to horizontally polarized input light scattered by polystyrene micro-sphere solution with $\mu_{s}' $ = 10 mm\textsuperscript{-1} and particle size $d = 0.1\  \mu m$, which represents an isotropic scatterer. The beams having non-zero OAM can be seen to have higher DOP than the fundamental Gaussian mode ($l$ = 0). 
To better observe the trend in the depolarization, we consider the variation of the DOP along a segment parallel to the y-axis passing through the center of the 2D plots in Fig.\ref{fig:particle_a} (a)-(d). This variation is plotted as a function of the distance from the beam center. Fig.\ref{fig:particle_a}(e) shows these plots in the case of $l$ = 0, 3, 6, 9 for input horizontal polarization. We observe that the measured DOP gets progressively higher with the increase in the input OAM (topological charge $l$) value. The slight asymmetry in the peak values on the two sides of the beam center is due to the alignment of the SLM.

Fig.\ref{fig:particle_a}(f) shows the measured DOP values along the selected segment for the isotropic scatterer in the case of left circularly polarized incident light. As in the case of the horizontally polarized incident light, the beam carrying non-zero OAM  shows higher DOP than the standard Gaussian mode ($l$ = 0). Further, an increase in the input OAM value again results in an increase in the measured DOP. Comparing Fig.\ref{fig:particle_a}(e), and (f), we observe that the measured values of DOP for a beam with a particular OAM value are higher in the case of horizontally polarized incident light as compared to circularly polarized incident light, including the standard Gaussian beam. This implies that circularly polarized light's depolarization is stronger than linearly polarized light for the forward scattered light collected from the isotropic scattering medium.


\begin{figure}[H]
\centering
\includegraphics[width=.75\linewidth]{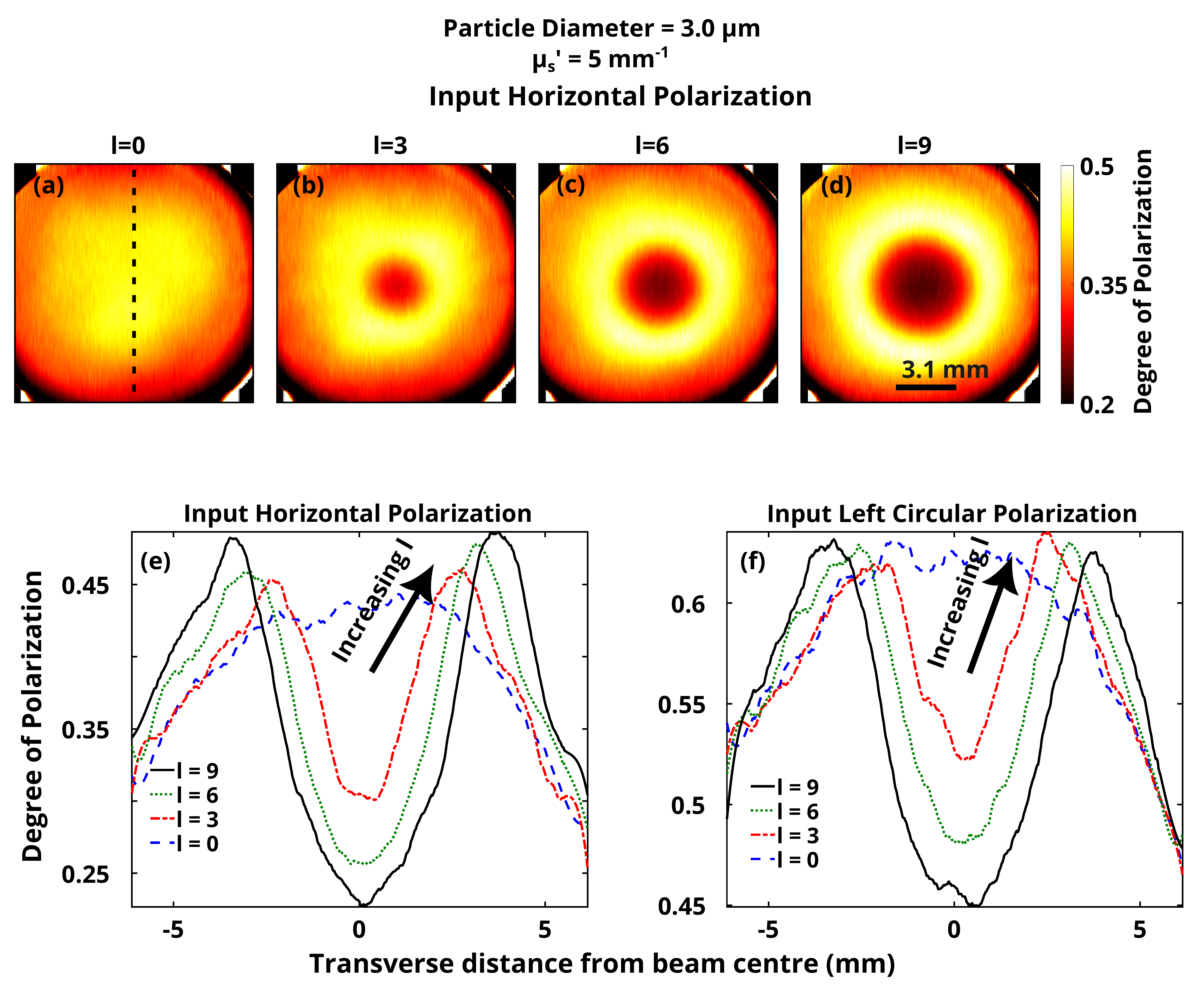}
\caption{\textbf{Influence of the OAM topological charge $l$ on the propagation of polarized light and the resultant degree of polarization of the transmitted light in anisotropic scattering media $(g=0.814, d=3\ \mu m, \mu_{s}'=5 mm^{-1})$}.\textbf{((a)-(d))} Measured pixel-wise DOP in transmitted LG beam for incident beam carrying topological charge $l$ = 0, 3, 6, 9 respectively. The colourbar represents the value of degree of polarization.  \textbf{(e)} and \textbf{(f)} DOP variation along a segment selected parallel to the y-axis in the pixel-wise DOP plot for incident horizontally and left circularly polarized light respectively. Representative segment is shown in  \textbf{(a)} .  DOP values are shown for $l$ = 0 (Blue dashed line), $l$ = 3 (Red dash-dotted line), $l$ = 6 (Green dotted line), $l$ = 9 (Black solid line). An increase in the $l$ value of the input light is seen to result in stronger propogation and retention of polarization of both linearly and circularly polarized beams. For a fixed value of $l$, depolarization of horizontally polarized light is observed to be stronger than circularly polarized light .} 
\label{fig: particle_b}
\end{figure}

We next examine the effect of the particle size by considering a turbid medium comprised of scatterers with particle size d= 3 $\mu m$. This corresponds to an anisotropic scatterer in the Mie scattering regime. As before, we measure the DOP of the transmitted beam after propagation  
 through the turbid medium. Fig.\ref{fig: particle_b}(a)-(d) shows the measured pixel-wise DOP as 2D plots for horizontally polarized light with $l$ = 0, 3, 6, 9 respectively for a polystyrene microsphere solution with  $\mu_{s}' $ = 5 mm\textsuperscript{-1}. Unlike the starkly varying DOP in the case of the isotropic scatterer, the measured DOP shows only a gradual increase with the OAM value in this case. In a similar fashion as before, we select a segment of the 2D plot parallel to the y-axis  and plot the DOP variation along this segment as a function of the distance from the beam centre. Fig.\ref{fig: particle_b}(e) shows this plot for the horizontally polarized input light. We observe that the dependence of the DOP on the input OAM closely resembles that which is previously seen in the case of the isotropic scattering media (Fig.\ref{fig:particle_a}). Indeed, the higher OAM modes display progressively higher survival in terms of their depolarization. Hence the increased survival of higher-order OAM beams is observed for both isotropic and anisotropic scattering media. 
 
Fig.\ref{fig: particle_b}(f) shows the measured DOP values for incident left circularly polarized light transmitted through the anisotropic scattering media. As consistently observed before, the measured DOP values increase with an increase in the OAM carried by the input beam. However, the relative rate of depolarization of the circularly polarized and linearly polarized light is contrary to what was observed in the case of the isotropic scattering media. For a given OAM value, the measured DOP values for horizontally polarized light (Fig.\ref{fig: particle_b}(e)) are lower than those measured for the left circularly polarized light (Fig.\ref{fig: particle_b}(f)). Hence the forward scattered light from the anisotropic scattering media reveals a weaker depolarization of the circularly polarized light as compared to the linearly polarized light.


\begin{figure}[H]
\centering
\includegraphics[width=.75\linewidth]{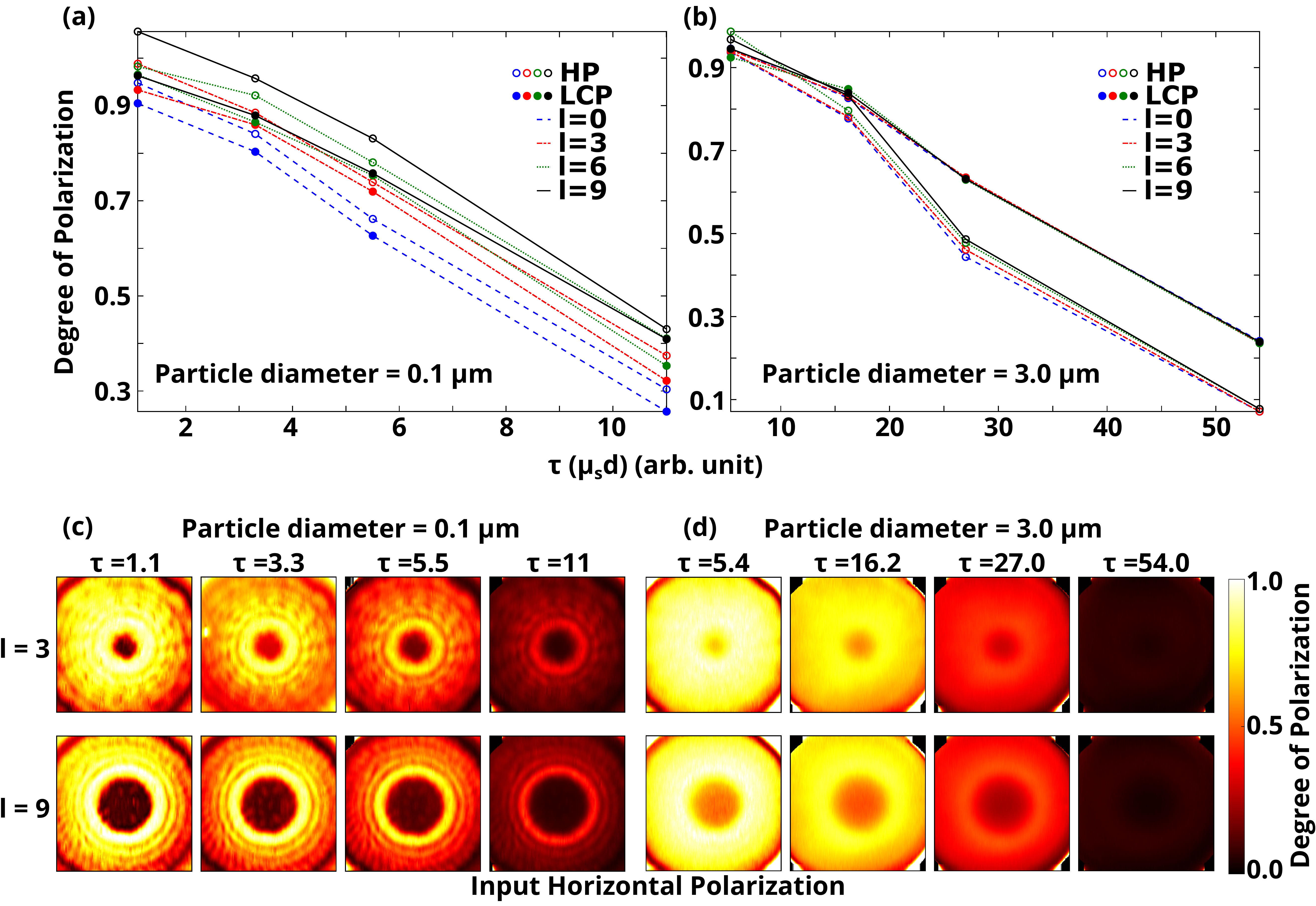}
\caption{\textbf{Dependence of the survival of the Lauguerre-Gaussian (LG) vortex beam and the depolarization of linearly and circularly polarized light on the optical thickness of both isotropic $(g = 0.091, d = 0.1\ \mu m)$ and anisotropic $(g = 0.814,  d = 3 \mu m)$ scattering media.} 
\textbf{(a)}, \textbf{(b)} Measured peak degree of polarization as a function of the optical thickness (\texttau) for a LG beam transmitted through isotropic and anisotropic scattering media respectively. DOP values are shown for $l$ = 0 (Blue dashed line), $l$ = 3 (Red dash-dotted line), $l$ = 6 (Green dotted line), $l$ = 9 (Black solid line). Input horizontal and circularly polarized light are represented by open-circle and closed-circle markers respectively. \textbf{(c)} Measured pixel-wise DOP as a function of optical thickness and particle size for input $l$ =3 and $l$ =  9. The colourbar represents the degree of polarization.}
\label{fig:concentration_dependence}
\end{figure}

Next, we investigate the effect of the optical thickness on the measured DOP values for the LG beam, comparing across the range of OAM values. The optical thickness $(\tau)$, defined as $\tau = \mu_{s} \times d$, is a unitless quantity and is a measure of the scattering efficacy of a turbid scattering media. For each particle size, we prepare samples with a fixed set of $\mu_{s}'$ values, as listed in Table \ref{tab:parameter_table}. Then we perform the polarization measurements to obtain the pixel-wise DOP plots as before. Selecting a segment parallel to the y-axis of the 2D plot, we obtain the DOP variation as a transverse distance from the beam centre. From this variation we obtain the peak value of the DOP and denote it as a  representative value for the survival of the beam.  
Fig.\ref{fig:concentration_dependence}(a) and (b) show the measured peak DOP values as a function of the optical thickness for particle size $d= 0.1\ \mu m$ and $d= 3\ \mu m$ respectively. As expected, the DOP values decrease with the increasing optical depth. Additionally, a stark contrast is observed in survival of the beam for the two particle sizes. While the measured DOP for the isotropic scatterer $(d = 0.1\ \mu m)$ falls below $0.5$ over an optical depth of $\tau = 11.006$, an optical depth of at least $27.007$ is needed for the anisotropic scatterer $(d = 3\ \mu m)$. Further, a marked difference in the rate of depolarization for circularly polarized and linearly polarized incident light is evident for each of the optical thickness values under consideration. The relative rates of depolarization are in agreement with the previously observed trends, namely circularly polarized light shows greater survival than linearly polarized light in the case of anisotropic scattering media, while the opposite is true in the case of isotropic scattering media.

The influence of the OAM value on the measured DOP 
has already been shown in the previous sections. Here we see that the extent of this influence is different in the two different kinds of scattering media. In fact, the effect of the changing OAM is more pronounced in the isotropic scattering media $(d= 0.1\ \mu m)$ than in the anisotropic scattering media $(d= 3\ \mu m)$. This can be easily seen from the difference in the measured DOP values for $l$ = 9 and $l$ = 0, given by the black solid line and blue dashed line respectively. For both linearly and circularly polarized light, these lines show a close match in the anisotropic scattering media (Fig.\ref{fig:concentration_dependence} (b)) while a significant difference in the DOP is observed in the isotropic scattering media(Fig.\ref{fig:concentration_dependence}(a)).

Fig.\ref{fig:concentration_dependence}(c), and (d) show the measured DOP as a function of the optical thickness as 2D plots, for particle size $d= 0.1\ \mu m$ and $d= 3\ \mu m$ respectively. Measured pixel-wise DOP are shown for $l$ = 3 and $l$ = 9 in the case of the horizontally polarized input light. We observe that the decrease in DOP with increase in the optical thickness occurs uniformly throughout the beam profile. We again observe that the survival of the beam is seen upto a higher optical thickness in the case of the anisotropic scatterer (Fig.\ref{fig:concentration_dependence}(d) than for the isotropic scatterer(Fig.\ref{fig:concentration_dependence}(c).


\begin{figure}[H]
\centering
\includegraphics[width=.75\linewidth]{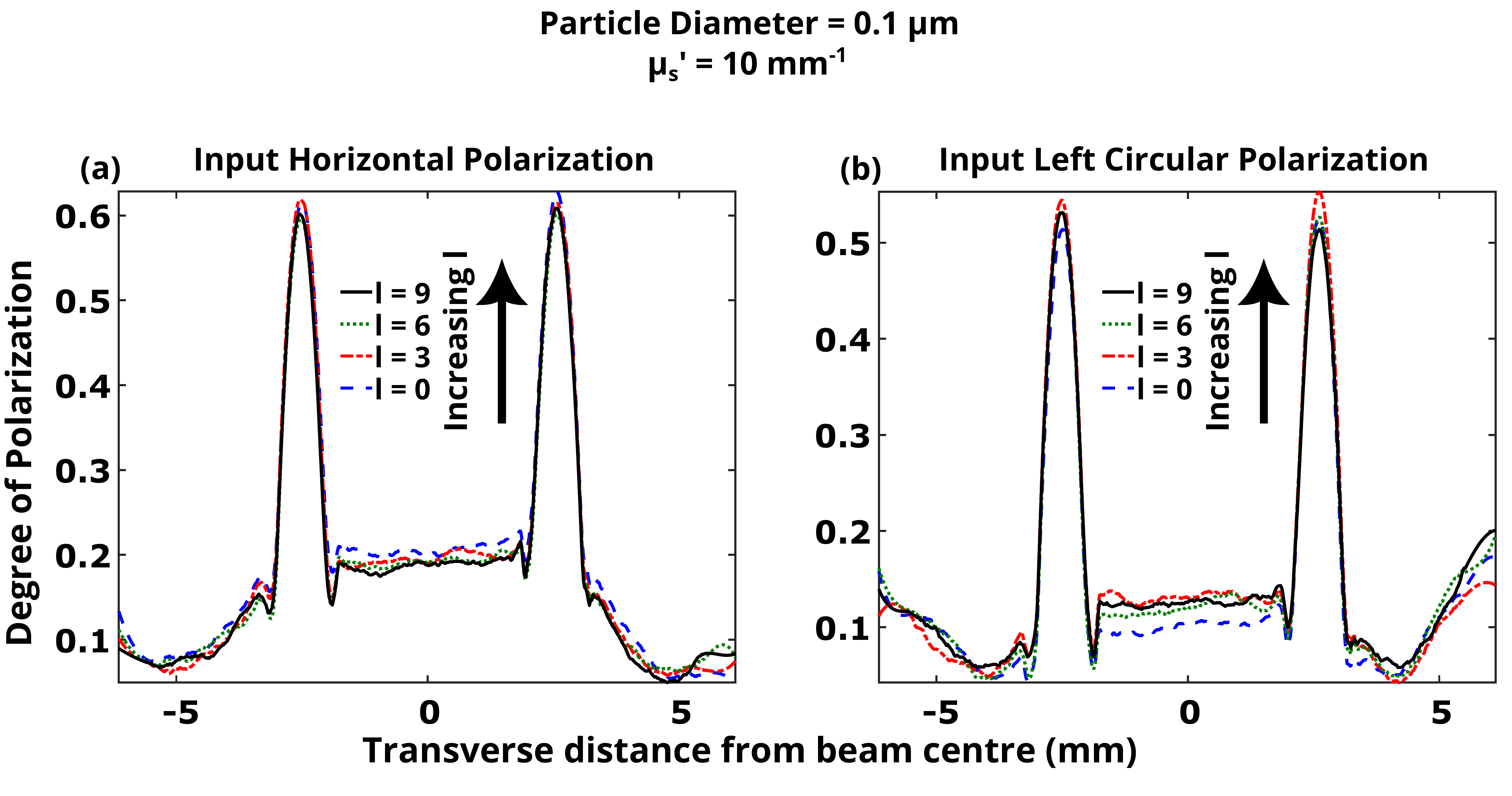}
\caption{\textbf{Dependence of the survival of the perfect vortex beam and the survival of polarization in isotropic scattering media $(g = 0.091, d= 0.1\ \mu m , \mu_{s}' = 10\  mm^{-1})$}  \textbf{(a)} and \textbf{(b)} DOP variation along a segment selected parallel to the y-axis in the pixel wise DOP plot for incident horizontally and left circularly polarized perfect vortex beam respectively. The DOP values are shown for $l$ = 0 (Blue dashed line), $l$ = 3 (Red dash-dotted line), $l$ = 6 (Green dotted line), $l$ = 9 (Black solid line). The change in input OAM results in no appreciable change in the measured DOP values in the transmitted beam. 
}
\label{fig:perfect_vortex}
\end{figure}

Finally, we perform DOP measurements by changing the LG beams into perfect vortex beams of different orders. The combination of the axicon and  the biconvex lens converts an input LG beam with topological charge $l$ into a perfect vortex beam with the same mode number $l$. The generated beam thus has an azimuthal phase gradient identical to the LG beam that was used to generate it. 
However the intensity profile remains constant for all the $l$ values including $l$ = 0. Fig.\ref{fig:perfect_vortex}(a) shows the variation of the DOP values in the transmitted perfect vortex beam along a segment selected parallel to the y-axis in the pixel-wise DOP plots. The DOP values are shown for $l$ = 0, 3, 6, 9 for horizontally polarized light incident on isotropic scattering media. We observe that the effect of the changing OAM value on the measured DOP is strikingly different from the observed effects in case of polarized LG beams. Particularly, the incremental rise of the measured DOP with the increase in OAM value is absent in the case of perfect vortex beams. In fact, the change in the DOP with changing OAM is severely diminished and is within the limits of error for the experiment. It is also evident that the peak values of the DOP occur at the same transverse distance from the beam center. This originates from the fact that the intensity profile is constant with the changing OAM values. 
Fig.\ref{fig:perfect_vortex} (b) shows the DOP variation along the selected segment for left circularly polarized incident light. No significant or progressive increase in the DOP values are observed in this case as well. Comparing Fig.\ref{fig:perfect_vortex}(a) and Fig.\ref{fig:perfect_vortex}(b), we observe that the measured DOP values for circularly polarized light are lower than the measured values for linearly polarized light for a given $l$ value.  Therefore the relative rates of depolarization for the two different polarization displays the same relationship as that of a LG beam scattered by isotropic scattering media.


The mechanism of depolarization of light through turbid media has been understood through the randomization of its propagation direction and polarization in the multiple scattering regime. The randomization of the propagation direction depends on the scattering anisotropy parameter (\textit{g}), which captures the dominant scattering directions for each scattering event. The \textit{g} parameter is in turn determined by the nature of the light-matter interaction and the properties of the scattering particle such as the refractive index, size and scattering cross-section. For large values of the \textit{g} parameter (\textit{g} $>$ 0.8), the scattering is preferentially in the forward direction. This results in lower direction randomization and hence stronger propagation through the scattering media. On the other hand, for lower \textit{g} values (\textit{g} $<$ 2), the scattering is isotropic, with a near-equal contribution to forward and backward scattering. Consequently, there is a greater randomization of the propagation direction, observed through a weaker propagation through scattering media. It is thus evident that the depolarization of light that experiences a higher \textit{g} parameter is lesser than light that experiences a low \textit{g} parameter. This explains the survival of the beam up to higher optical thickness values in the case of the anisotropic scattering media $(d = 3\ \mu m $, \textit{g} = 0.814) than for the isotropic scattering media $(d = 0.01\ \mu m $, \textit{g} = 0.091) as seen in by comparing Fig.\ref{fig:concentration_dependence}(a)and (b).

The dissimilar rates of depolarization of linearly and circularly polarized light have also been well understood in relation to the \textit{g} parameter. In addition to the direction randomization described above, circularly polarized light also has the possibility to undergo a change in its helicity through scattering. For the helicity flip to occur, a large scattering angle is necessary.
For a large value of the \textit{g} parameter, a smaller proportion of light is scattered at larger angles, as the preferred scattering is in the forward direction. This results in a weak helicity flip. In contrast, for small \textit{g} values, the helicity flip is stronger due to a greater proportion of light scattered at large angles. The effect of helicity flipping is thus significant in the isotropic scattering media and weak in anisotropic scattering media. This explains the stronger depolarization of circularly polarized light for the isotropic scattering media $(d = 0.01\ \mu m $, \textit{g}= 0.091) and weaker depolarization in anisotropic scattering media $(d = 3\ \mu m $, \textit{g} = 0.814). Thus the varying trends in depolarization again have a clear correlation with the g parameter.

The stronger propagation of the LG beams as compared to the standard Gaussian beams has been clearly established by the measured higher DOP. The major properties that set these two classes of beams apart are a) The azimuthally varying phase structure and b) the Mode-dependent ring shaped intensity profile of the LG beams. The stronger propagation of the LG beams can then be ascribed to either (a) and (b) or a combination of both. The measured depolarization of the perfect vortex beams reveals crucial information about the scattering behavior. The perfect vortex beams have associated with them an azimuthally varying phase profile, similar to LG beams. In fact, we have generated the perfect vortex beams by acting upon the LG beams with a combination of an axicon lens and a convex lens, which do not impart any azimuthally varying phase dependence. Clearly, if the stronger propagation of the beams has a contribution from the varying phase profile, it should manifest as a change in the measured DOP with changing mode number of the perfect vortex beam. However, no significant change in the DOP is observed in this case. In fact the $l$ = 0 mode also exhibits the same measured DOP as the higher-order modes. This observation indicates that the observed longer survival of higher OAM modes is not influenced by the true phase profile but is mainly related to the intensity profile of the beam. Hence the mode-dependent intensity profile of the LG beams can be said to govern the depolarization characteristics and contribute to the robust propagation properties. 

Since the stronger propagation is thought to arise from the intensity profile, it is important to address another effect associated with the changing intensity profile of the LG beams. It is known that the thickness of the LG beam ring decreases with the increasing ring radius and OAM value. Hence we note that beams with higher $l$ values present with slightly higher intensities on the ring, following the conservation of energy. However, the calculated stokes parameters are independent of the input intensity and hence the observed trend does not originate from the variation in the intensities. This is confirmed by the DOP values measured for the SDS solution (See Supplementary S.4), in which no trend in the DOP is observed in spite of varying input intensities. Hence we verify that the observed changing DOP values are due to the varying interaction of the various LG beam orders with the turbid medium. This observation is in agreement with the increased transmittance with higher OAM values as seen in other studies \cite{Deeptransmission,chickentissue}. 

As already indicated, the \textit{g} parameter plays a decisive role in the scattering behavior and has a clear correlation with the depolarization characteristics. The longer survival of the LG beams due to the intensity profile can be explained by a change in the effective anisotropy parameter \textit{g} as seen by the beam. Indeed, a higher experienced \textit{g} parameter with the increase in OAM value and hence a change in the associated intensity profile will result in stronger propagation of the beam through the scattering media. Further, our results have also shown that while overall propagation is stronger in the anisotropic scattering media (g $>$ 0.8), the effect of changing OAM is weaker(See Fig. \ref{fig:concentration_dependence}). In contrast, the effect of the changing intensity profile is relatively stronger in the isotropic scattering media. Hence the extent to which the change in the intensity changes the depolarization behavior also depends on the calculated g value from Mie theory. This observation further reinforces the presence of a link between the g parameter and the observed scattering differences of the LG beams. This also implies that the effect of the changing intensity profile is more pronounced in the regime which displays greater depolarization,  precisely where the need for stronger propagation is higher.

\section{Conclusion}
\label{section4}
In summary, we have performed degree of polarization measurements for the forward scattered light in the case of both LG beams and perfect vortex beams with varying topological charge \textit{l}. A comparison between the observed depolarization trends for the two classes of OAM-carrying beams suggests that the robust scattering properties of the LG beams originate from the intensity profile while the phase profile does not seem to play a major role in the scattering process. This insight can guide future research in engineering of novel states of light whose intensity profiles offer resilience in environments dominated by scattering. Additionally, the stronger survival of the higher order LG beams originates from the higher scattering anisotropy factor \textit{g} experienced. Further studies are required in order to describe the effects of an arbitrary intensity profile on the effective g parameter. 
We hope our studies inspire further investigations towards the development of a concrete model accounting for the behavior of structured light in scattering media. With this knowledge it will be possible to generate beams which will be resilient to scattering in the multiple scattering regime, allowing advancement of technologies for imaging, communication and sensing applications.

\section{Author Contributions}
\label{author}
SG \& AP contributed equally to this work.

\section{Data Availability Statement}
\label{dataavail}
The data that support the findings of this study are available upon reasonable request from the authors.

\section{Acknowledgement}
\label{acknow}
The authors thank the support of Indian Institute of Science Education and Research Kolkata(IISER-K), Ministry of Education, Government of India. The authors would like to acknowledge the Science and Engineering Research Board (SERB), Government of India, for the funding (grant No. CRG/2019/005558). SG additionally acknowledges CSIR, Government of India, for research fellowships. We like to acknowledge Niladri Modak for his help in building the experimental setup for collecting diffuse light.

\section{Conflict of Interest}
\label{conflict}
The authors declare no conflict of interest.

\

\printbibliography

\end{document}


\pagenumbering{arabic}
\title{Exploring the origin of stronger survival of polarized vortex beams through scattering media}
\author[1*]{Atharva Paranjape}

\author[1*+]{Shyamal Guchhait}

\author[2]{Athira B S}

\author[1,3]{Nirmalya Ghosh}
\affil[1]{Department of Physical Sciences\\

Indian Institute of Science Education and Research Kolkata\\

Mohanpur, India - 741246}
\affil[2]{Department of Bioengineering\\
University of Washington\\
Seattle, USA - 98105}

\affil[3]{Centre of Excellence in Space Sciences India\\

Indian Institute of Science Education and Research Kolkata\\

Mohanpur, India- 741246}

\affil[+]{\href{mailto:sg16ip022@iiserkol.ac.in}{sg16ip022@iiserkol.ac.in}}
\affil[*]{these authors have contributed equally to this work}

\date{}
\maketitle
\renewcommand\thesection{S}
\section{Supplementary information}
\subsection{LG beam generation}

\begin{figure}[ht]
    \centering
    \includegraphics[width=1\linewidth]{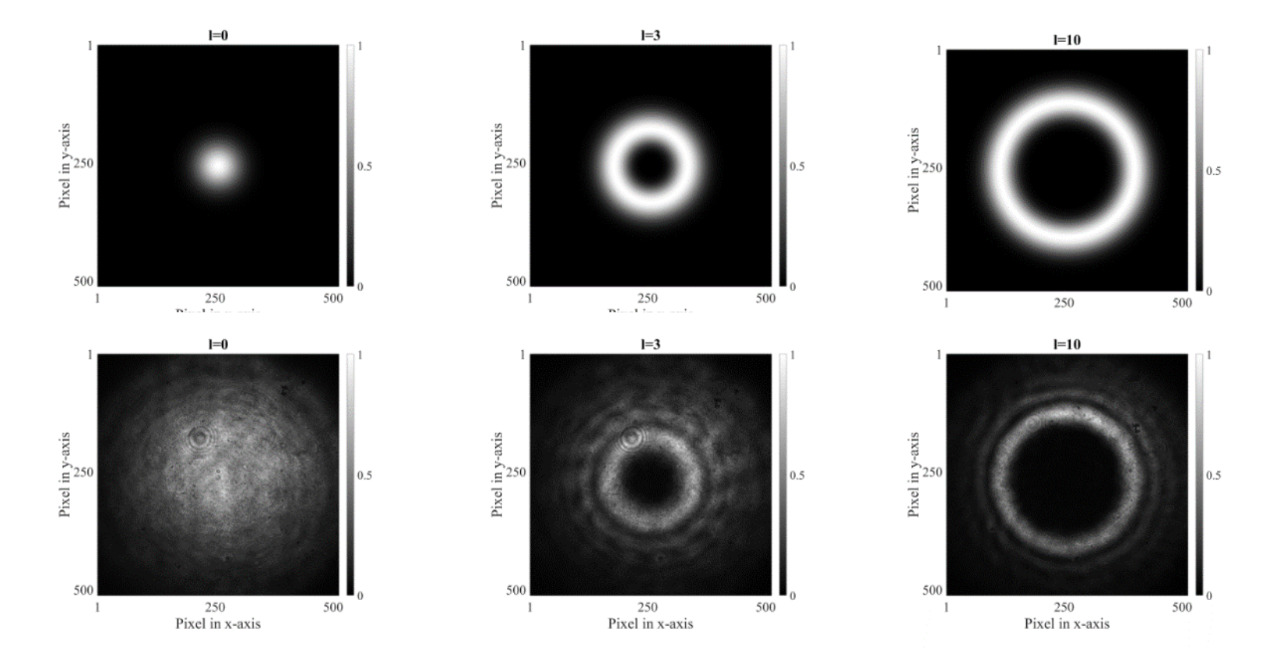}
    \caption{ Theoretical vs experimental intensity distributions of Laguerre-Gaussian beams with topological charge \emph{l} = 0, 3 and 10 as used in the experiment. The colourbar represents intensity in arb. units}
    \label{LG_generation}
\end{figure}

Our experiment strongly depends on the reliable generation of the Laguerre-Gaussian (LG) beams. Figure \ref{LG_generation}
 shows the unscattered input light compared with theoretical intensity plots of the LG beams for \emph{l}= 0, 3 and 10. We observe a close match between the theoretical and experimental intensity patterns indicating attested generation of the LG beams.  

\subsection{Sample preparation}

The scattering effectiveness 
of the microsphere solution is captured by the scattering coefficient $\mu_{s}$ = $\sigma_{s}$ $\cross$ N , where $\sigma_{s}$  is the scattering cross section and N is the number of particles per unit volume. In the case of polysterene microspheres, the absorption coefficient is zero, and the scattering coefficient ($\mu_{s}$) is equal to the extinction coefficient ($\mu_{e}$) of the solution. The reduced scattering coefficient is given by $\mu_{s}'$ = $\mu_{s}$ $\cross$ (1 - g) where g is the scattering anisotropy parameter. $\mu_{s}'$  thus captures the scattering effectiveness as well as the dominant scattering direction exhibited by the particles in the solution. For given values of the wavelength of light, particle diameter(d), concentration (N), refractive indices of the scatterer and solvent, Mie theory provides the $\mu_{s}'$, $\mu_{s}$ and g  values. 

To investigate the multiple scattering as well as ballistic scattering regime,  we select a particular set of \textmu\textsubscript{s}\textsuperscript{'} values (10 mm\textsuperscript{-1}, 5 mm\textsuperscript{-1}, 3 mm\textsuperscript{-1}, 1 mm\textsuperscript{-1} ). For each particle size,  the concentration (N) required to acquire the desired $\mu_{s}'$ is calculated using a code for Mie scattering developed in-house. SDS solution (0.1 \% w/v) is prepared using distilled water and powdered SDS from Sigma-Aldrich. Stock solutions of the microsphere beads (Sigma-Aldrich ) are then diluted using the SDS solution.  The same glass cuvette (path length = 1mm) is used throughout the experiment. The cuvette is cleaned with isopropanol and distilled water before insertion of a subsequent solution. During the experiment the cuvette is placed in an elevated sample holder which allows collection of the diffuse light.




\


\subsection{Diffuse light collection and image capture}

\begin{figure}[H]
    \centering
    \includegraphics[width=.75\linewidth]{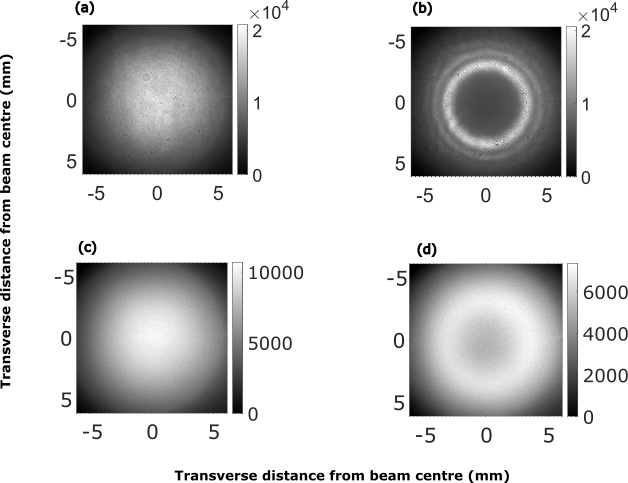}
    \caption{ \textbf{Collected light in co-polarized (A, B)  and cross-polarized (C ,D ) configurations.} Images correspond to input light with topological charge l=0 (A,C) and l=10(B,D) with linear polarization, for a sample particle diameter of 0.1 μm and scattering coefficient $\mu_s' = 10\ mm^{-1}$. The colourbar represents intensity in arb. units 
    }
    \label{fig:diffuse}
\end{figure}

For accurate polarimetry, collection of the diffuse scattered light from the sample is crucial. Fig. \ref{fig:diffuse} shows the intensity plots of the collected light in the forward direction for various relative positions of the polarizer (P1) and analyzer (P2). As we work with largely varying concentrations, the intensity of transmitted light varies accordingly.

To avoid the  saturation of the images captured by the camera, the integration time and the value of the neutral density(ND) filter are adjusted. Table \ref{tab:integrationtimes} shows these value for the Laguerre- Gaussian as well as perfect vortex beams as incident light.  For a particular particle size and \textmu \textsubscript{s}\textsuperscript{'} value, the neutral density filter is kept constant during the degree of polarization (DOP) measurements. Hence the input intensity remains constant for a given sample. In the case of the Laguerre-Gaussian beam , the integration time is set at  0.1 s irrespective of particle size or concentration. \ For the perfect vortex beams the integration time needs to be adjusted depending on whether the polarizer (P1) and analyzer (P2) are in a co-polarized or cross-polarized configuration. This is due to the limited dynamic range of the camera and high intensity of the perfect ring causing large intensity variation between co-polarized and cross-polarized configurations. For sample with particle size d = 0.1 $\mu$m,  All images in a cross-polarized configuration are captured with 2 s integration time, while images in a co-polarized configuration are captured with 0.05 s integration time.
Before the degree of polarization calculation, each image is normalized by dividing with the integration time, and the resulting values for unit time are used for DOP calculation. 
For sample with particle size d = 3 $\mu$m, the integration time is again set constant at 0.1 s.

 
\begin{table}[H]
\centering
\begin{tabular}{|c|c|c|c|c|}
\hline
Input Light                                                                       & \begin{tabular}[c]{@{}c@{}}$\mu_{s}'$\\ $(mm^{-1})$\end{tabular} & \begin{tabular}[c]{@{}c@{}}Particle \\ diameter(d)\\($\mu$m)\end{tabular} & \begin{tabular}[c]{@{}c@{}}Neutral Density\\ (ND filter value)\end{tabular} & \begin{tabular}[c]{@{}c@{}}Camera integration time\\  (s)\end{tabular} \\ \hline
\multirow{9}{*}{\begin{tabular}[c]{@{}c@{}}Laguerre-Gaussian\\ Beam\end{tabular}} & 10                                                               & 0.1                                                             & 2                                                                           & 0.1                                                                    \\ \cline{2-5} 
                                                                                  & 5                                                                & 0.1                                                             & 2.4                                                                         & 0.1                                                                    \\ \cline{2-5} 
                                                                                  & 3                                                                & 0.1                                                             & 3                                                                           & 0.1                                                                    \\ \cline{2-5} 
                                                                                  & 1                                                                & 0.1                                                             & 3.4                                                                         & 0.1                                                                    \\ \cline{2-5} 
                                                                                  &                                                                  &                                                                 &                                                                             &                                                                        \\ \cline{2-5} 
                                                                                  & 10                                                               & 3                                                               & 2                                                                           & 0.1                                                                    \\ \cline{2-5} 
                                                                                  & 5                                                                & 3                                                               & 2                                                                           & 0.1                                                                    \\ \cline{2-5} 
                                                                                  & 3                                                                & 3                                                               & 2                                                                           & 0.1                                                                    \\ \cline{2-5} 
                                                                                  & 1                                                                & 3                                                               & 3                                                                           & 0.1                                                                    \\ \hline
                                                                                  &                                                                  &                                                                 &                                                                             &                                                                        \\ \hline
\multirow{7}{*}{\begin{tabular}[c]{@{}c@{}}Perfect Vortex\\ Beam\end{tabular}}    & 10                                                               & 0.1                                                             & 2.5                                                                         & 0.05,2                                                                 \\ \cline{2-5} 
                                                                                  & 5                                                                & 0.1                                                             & 3                                                                           & 0.05,2                                                                 \\ \cline{2-5} 
                                                                                  & 3                                                                & 0.1                                                             & 4                                                                           & 0.05,2                                                                 \\ \cline{2-5} 
                                                                                  &                                                                  &                                                                 &                                                                             &                                                                        \\ \cline{2-5} 
                                                                                  & 10                                                               & 3                                                               & 1.5                                                                         & 0.1,0.1                                                                \\ \cline{2-5} 
                                                                                  & 5                                                                & 3                                                               & 2                                                                           & 0.1,0.1                                                                \\ \cline{2-5} 
                                                                                  & 3                                                                & 3                                                               & 2.5                                                                         & 0.1,0.1                                                                \\ \hline
\end{tabular}
\caption{Set of integration times and neutral density filter values used during degree of polarization measurements}
\label{tab:integrationtimes}
\end{table}

\subsection{Control sample DOP measurements}

\begin{figure}[H]
\centering
\includegraphics[width=.75\linewidth]{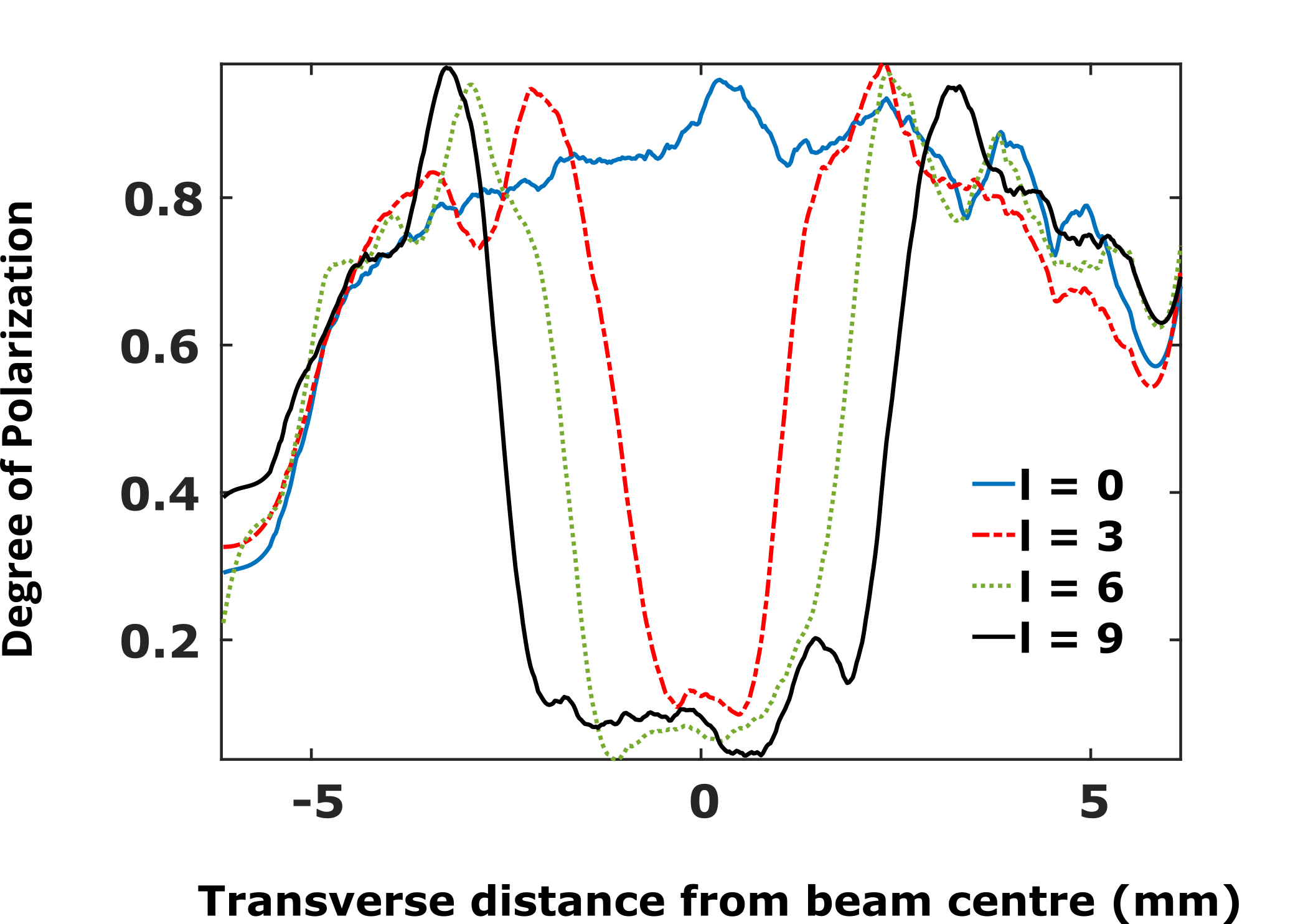}
\caption{\textbf{Influence of the OAM topological charge \emph{l} on the propagation of polarized light and the resultant degree of polarization of the transmitted light in SDS }. DOP variation along a segment selected parallel to the y-axis in the pixel-wise DOP plot for incident horizontally polarized light. DOP values are shown for \emph{l}=0 (Blue dashed line), \emph{l}=3 (Red dash-dotted line), \emph{l}=6 (Green dotted line), \emph{l}=9 (Black solid line) }
\label{fig:sds}
\end{figure}
To observe the effect of the varying \emph{l} on the measured DOP values, we use 0.1 $\%$ SDS solution in place of the sample. Following the procedure described in the main text, we measure the DOP for varying value of \emph{l} in the same experimental setup. Fig. (\ref{fig:sds}) show the measured DOP values in this case. We observe that the DOP does not show any progressive increase with an increase in the input OAM value.

